\documentclass[journal,10pt]{IEEEtran}

\IEEEoverridecommandlockouts 
\usepackage{epsfig,latexsym}
\usepackage{float}
\usepackage{indentfirst}
\usepackage{amsmath}
\usepackage{amssymb}
\usepackage{cite}
\usepackage{bm}
\usepackage{url}
\usepackage{lastpage}
\usepackage{color}
\usepackage{fancyhdr}
\usepackage{multirow}
\usepackage[square, comma, sort&compress, numbers]{natbib} 
\normalsize

\begin{document}


\title{Design of Protograph Codes for Additive White Symmetric Alpha-Stable Noise Channels}

\author{Xingwei Zhong,
Kui Cai,~\IEEEmembership{Senior Member,~IEEE}, Pingping Chen, and Zhen Mei
\thanks{ X. Zhong, K. Cai, P. Chen, and Z. Mei are with the Department of Science, Singapore University of Technology and Design, Singapore. (e-mail: xingwei$\_$zhong@mymail.sutd.edu.sg; cai$\_$kui@sutd.edu.sg; mei$\_$zhen@sutd.edu.sg).} 
\thanks{P. Chen is also with the College of Physics and Information, Fuzhou University, China. (e-mail: ppchen.xm@gmail.com)}
}

\maketitle

\begin{abstract}
The protograph low-density parity-check (LDPC) codes possess many attractive properties, such as the low encoding/decoding complexity and better error floor performance, and hence have been successfully applied to different types of communication and data storage channels. In this paper, we design protograph LDPC codes for communication systems corrupted by the impulsive noise, which are modeled as additive white symmetric alpha-stable noise (AWS$\alpha$SN) channels. We start by presenting a novel simulation-based protograph extrinsic information transfer (P-EXIT) analysis to derive the iterative decoding threshold of the protograph codes. By further applying the asymptotic weight distribution (AWD) analysis, we design new protograph codes for the AWS$\alpha$SN channel. Both theoretical analysis and simulation results demonstrate that the proposed protograph codes can provide better error rate performance than the prior art AR4JA code, the irregular codes optimized for the AWGN channel, as well as the irregular codes optimized for the AWS$\alpha$SN channel.
\end{abstract}

\section{Introduction}\label{sec1}

Many communication systems are corrupted by the impulsive noise, such as the indoor wireless communication system \cite{blackard1993measurements}, the shallow water acoustic system \cite{qarabaqi2009statistical}, and the power-line communication system \cite{laguna2015use,middleton1999non}. The impulse noise is non-Gaussian distributed, whose first order probability density function (PDF) follows the symmetric alpha-stable (S$\alpha$S) law \cite{middleton1999non}. Therefore, the corresponding communication systems can be modeled as additive white symmetric alpha-stable noise (AWS$\alpha$SN) channels.

In the literature, the low-density parity-check (LDPC) codes have been studied for the AWS$\alpha$SN channel as a type of capacity-approaching channel codes. In particular, their asymptotic performance with optimal and sub-optimal $\log$-likelihood ratios (LLRs) is investigated by \cite{maad2010asymptotic} through density evolution (DE). The waterfall region performance of LDPC codes for the AWS$\alpha$SN channel is analyzed by \cite{mei2016finite,mei2017performance}. Recently, a quantized DE (QDE) based extrinsic information transfer (EXIT) chart is applied to construct capacity-approaching code ensembles for the AWS$\alpha$SN channel \cite{dai2017exit}. However, as the parity-check matrices have irregular structures, the proposed code ensembles are not suitable for practical implementations.

In the recent few years, the protograph LDPC codes have been found to achieve superior error performance over the additive white Gaussian noise (AWGN) channel, the partial response (PR) channels, and the high-density magnetic recording channels \cite{liva2007protograph,divsalar2009capacity,chen2015design,zhong2018rate}. Furthermore, their protograph structure facilitates low-complexity encoding with readily parallelizable decoder implementations. For example, the AR4JA codes are well-known capacity-approaching protograph codes optimized for the AWGN channels \cite{divsalar2009capacity}. However, up till now, no work has been carried out for designing protograph LDPC codes for the AWS$\alpha$SN channel.

As the alpha-stable noise is non-Gaussian, the conventional closed-form protograph EXIT (P-EXIT) analysis \cite{liva2007protograph} derived for the AWGN channel cannot be directly applied to the AWS$\alpha$SN channel. Although a QDE based EXIT analysis \cite{dai2017exit} can be extended to measure the decoding threshold of the protograph codes with the help of full-DE \cite[\textbf{7.2.2}]{richardson2008modern} for the AWS$\alpha$SN channel, it will have huge computational complexity \cite{richardson2008modern}, and hence is impractical. In this paper, we present a novel simulation-based P-EXIT analysis to compute and optimize the decoding threshold of the protograph LDPC codes. Through further analyzing the asymptotic weight distribution (AWD) of the code ensembles, we construct new protograph LDPC codes which achieve better error rate performance than the prior art AR4JA code, the irregular codes optimized for the AWGN channel, as well as the irregular codes optimized for the AWS$\alpha$SN channel \cite{dai2017exit}.

In the rest of the paper, Section 2 gives a review of the AWS$\alpha$SN channel model. Section 3 presents a novel simulation-based P-EXIT analysis for the AWS$\alpha$SN channel. In Section 4, we design new protograph LDPC codes for the AWS$\alpha$SN channel. Section 5 compares their error rate performance with the prior art LDPC codes. The paper is concluded by Section 6.

\begin{figure*}[t]
\centering
\includegraphics[height=0.8\columnwidth]{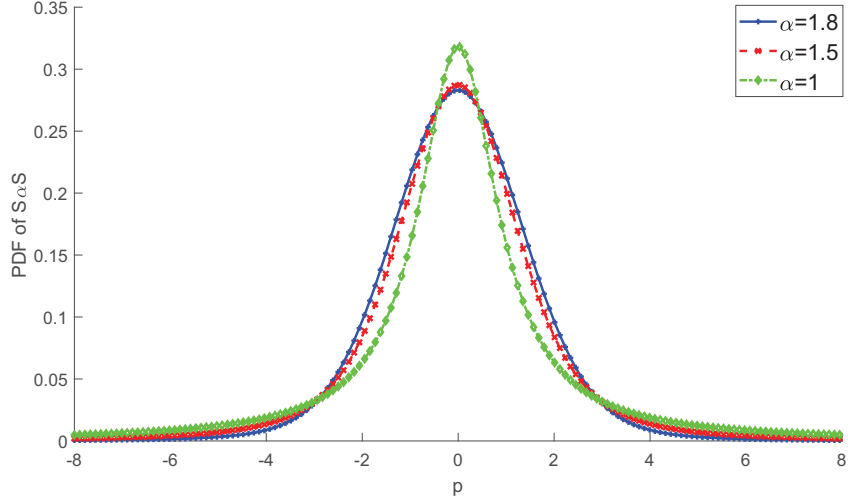}
\caption{PDFs of standard S$\alpha$S distributions with different $\alpha$ values (with $\gamma = 1$).}
\label{pdf_figure}
\end{figure*}

\section{Channel Model}

We adopt a characteristic function to define the S$\alpha$S distribution, given by \cite{dai2017exit}:
\begin{equation} \label{symmetric}
\phi(l)=e^{(-\gamma^{\alpha} |l|^{\alpha})},
\end{equation}
where $\gamma>0$ is the dispersion that measures the spread of the PDF. Here, $\alpha$, ranging from 0 to 2, is the characteristic exponent which sets the degree of impulsiveness of the PDF. When $\alpha=2$, the noise becomes Gaussian and hence will not be considered in this work. When $\alpha=1$, the noise is actually Cauchy distributed. For other values of $\alpha$, the noise follows the S$\alpha$S distribution and will have no closed-form expression. With the decrease of $\alpha$, the degree of impulsiveness of the distribution increases. Therefore, the PDF of an S$\alpha$S random variable is given by
\begin{equation} \label{pdf}
f_{\alpha}(p; \gamma)=\frac{1}{2\pi}\int_{-\infty}^{\infty}\exp(-\gamma^{\alpha} |l|^{\alpha})e^{-jlp}dl .
\end{equation}
We illustrate by Fig. \ref{pdf_figure} the PDFs of S$\alpha$S distributions with different values of $\alpha$, for $\gamma = 1$.

Furthermore, the additive noise model can be expressed as
\begin{equation} \label{model}
F=P+B,
\end{equation}
where $F$ is the channel output signal, $P \in \{-1, 1\}$ is the channel input signal with binary phase-shift keying (BPSK) constellation, and $B$ is the alpha-stable channel noise. We follow \cite{gonzalez2006zero} and adopt the geometric signal-to-noise ratio (G-SNR) defined for the AWS$\alpha$SN channel, given by
\begin{equation} \label{GSNR}
\text{G-SNR} = \frac{1}{2C_{g}^{(\frac{2}{\alpha}-1)}\gamma^2},
\end{equation}
where $C_{g}\approx 1.78$ denotes the Euler's exponential constant. The corresponding $\frac{E_{b}}{N_{0}}$ is calculated as
\begin{equation} \label{ebn0}
\frac{E_{b}}{N_{0}} = \frac{\text{G-SNR}}{2R},
\end{equation}
where $R$ is the code rate.

\section{Analysis of Protograph Codes for the AWS$\alpha$SN Channel}
A protograph can be represented by a small tanner graph consisting $T$ check nodes (CNs), $S$ variable nodes (VNs), and the edges which connect the CNs with the VNs. Unlike other types of LDPC codes, a protograph can contain parallel edges. We refer a $S \times T$ adjacency matrix $\pmb{B}$ corresponding to a protograph as the base matrix, with $b_{st}$ being the $(s,t)^{th}$ entry of $\pmb{B}$. For example, the base matrix of the AR4JA protograph code which shows superior performance over the AWGN channel \cite{divsalar2009capacity}, is given by
\begin{equation}  \label{basematrixB}
 \pmb{B}_{A4}=\left(
  \begin{array}{ccccc}
    1 & 2 & 0 & 0 & 0\\
    0 & 3 & 1 & 1 & 1\\
    0 & 1 & 2 & 2 & 1\\
  \end{array}
\right),
\end{equation}
where the corresponding protograph consists of 5 VNs, 3 CNs, and 15 edges. Through a sequence of ``copy-and-permute'' operations on $\pmb{B}$, the tanner graph of an LDPC code with any desired block length can be derived \cite{liva2007protograph}. During the design of protograph LDPC codes, the P-EXIT analysis \cite{liva2007protograph,chen2015design}, together with the AWD analysis \cite{divsalar2009capacity}, are effective theoretical methods to evaluate the decoding performance of the designed protograph code at the waterfall region and error floor region, respectively. 


\subsection{P-EXIT Analysis for the AWGN Channel}

For the AWGN channel, the P-EXIT analysis can track the convergence behavior of iterative decoding based on a closed-form multi-dimensional technique \cite{liva2007protograph,ten2004design}. Let $J(\sigma)$ denote the mutual information (MI) $I$ between the LLR value $L_{ch}$ and a coded bit \cite{ten2004design}. we have

\begin{equation} \label{Jfunction}
J(\sigma) = 1-\int_{-\infty}^{\infty}\frac{E(-((t-\sigma^2/2)^2/(2\sigma^2)))}{\sqrt{(2\pi\sigma^2)}}*log_{2}[1+E(-t)]dt 
\end{equation}  


With the help of the function $J$ and its inverse $J^{-1}$, the P-EXIT analysis proposed for the AWGN channel \cite{liva2007protograph} can be summarized as follows: \\
1) {\bf Initialization}. Select a $E_{b}/N_{0}$. Initializing a vector $\pmb{\sigma_{ch}}=(\sigma_{ch,0},\ldots,\sigma_{ch,T-1})$ such that:
\begin{equation} \label{sigmach}
\sigma_{ch,t}^2=8R(E_{b}/N_{0})_{t}.
\end{equation}
where $R$ is the code rate of the protograph. If $t$ is the puncture node, set $\sigma_{ch,t}=0$. \\
2) {\bf VN to CN Renewal}. For $t\in \lbrace 0, \ldots ,T-1 \rbrace$ and $s\in \lbrace 0, \ldots ,S-1 \rbrace$, compute
\begin{equation} \label{LEij1}
I^{(s,t)}_{E,V}=J\left(\sqrt{\sigma_{ch,t}^2+\sum_{c=1}^{S}(b_{ct}-\delta_{cs})(J^{-1}(I^{(c,t)}_{E,C}))^2}\right).
\end{equation}
If $c=s$, set $\delta_{cs}=1$. Otherwise, set $\delta_{cs}=0$. \\
3) {\bf CN to VN Renewal}. For $t\in \lbrace 0, \ldots ,T-1 \rbrace$ and $s\in \lbrace 0, \ldots ,S-1 \rbrace$, compute

\begin{equation} \label{LCij1}
I^{(s,t)}_{E,C}=1-J\left( \sqrt{\sum_{x=1}^{T}(b_{sx}-\delta_{xt})
(J^{-1}(1-I^{(s,x)}_{E,V}))^2}\right) .
\end{equation}
If $x=t$, set $\delta_{xt}=1$. Otherwise, set $\delta_{xt}=0$.  \\
4) {\bf A Posteriori MI Accumulation}. For $t\in \lbrace 0, \ldots ,T-1 \rbrace$, compute
\begin{equation} \label{Lmie1}
I^{t}_{MIE}=J\left(\sqrt{\sigma_{ch,t}^2+\sum_{c=1}^{S}(J^{-1}(I^{(c,t)}_{E,C}))^2}\right).
\end{equation}
\\
5) {\bf Stopping criterion}. Return to Step 2 and carry on the iteration until $I^{t}_{MIE}=1$ for all $t$. The decoding threshold is the lowest value $E_{b}/N_{0}$ that meets the stopping criterion.

However, the above P-EXIT analysis derived for the AWGN channel cannot be applied to the AWS$\alpha$SN channel since the S$\alpha$S noise is non-Gaussian. The QDE based EXIT analysis is also impractical for the AWS$\alpha$SN channel due to its huge computational complexity for the desired protograph codes. Hence, in the following, a simulation-based EXIT function is first derived for the AWS$\alpha$SN channel, based on which a novel P-EXIT analysis is proposed for the AWS$\alpha$SN channel.

\begin{figure*}[t]
\centering
\includegraphics[height=0.9\columnwidth]{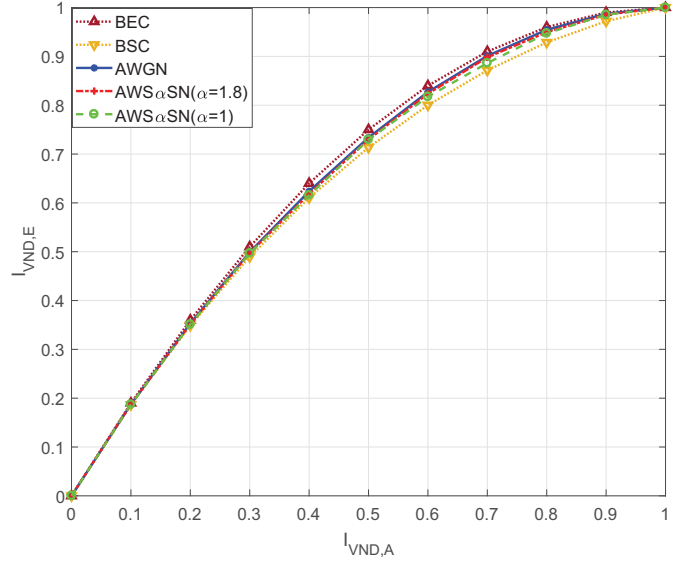}
\caption{VND EXIT curves for length 3 repetition codes over the BEC, BSC, AWGN,  and AWS$\alpha$SN channels (with $\alpha=1.8$ and $\alpha=1$).}
\label{IaverursIe}
\end{figure*}

\subsection{EXIT Function for the AWS$\alpha$SN Channel}

The EXIT analysis is designed to track the LLRs exchanged between the CN decoder (CND) and the VN decoder (VND). The LLR output from the VND is computed by \cite{ten2004design}:
\begin{equation}
L_{out,i}=L_{ch}+\sum_{j\ne i}^{}L_{in,j},
\end{equation}
with $i=1,2,\ldots,d_{v}$, and $d_{v}$ being the degree of the VN. Here, $L_{out,i}$ represents the $i$th output extrinsic LLR of the VN, $L_{in,j}$ denotes the $j$th {\it a priori} LLR, and $L_{ch}$ is the channel LLR. Unlike the AWGN channel, the PDF of the S$\alpha$S distribution cannot be described in the closed form. Therefore, we apply the numerical method \cite{nolan1997numerical} to compute $P(F|P)$, the conditional PDF of the AWS$\alpha$SN channel, as well as the corresponding channel LLRs given by $L_{ch}=\log\frac{P(F|P=+1)}{P(F|P=-1)}$.

On the other hand, as it is difficult to derive the $L_{in}$ of the AWS$\alpha$SN channel analytically, we propose the following approximation method. It has been found that the EXIT curves for the binary symmetric channel (BSC) and BEC provide the lower and upper bounds, respectively, for repetition codes over the binary-input, symmetric channel (BISC) \cite{land2005bounds,ashikhmin2004extrinsic}. Since both the AWGN channel and the AWS$\alpha$SN channel are BISCs, the EXIT curves for both the channels should lie between the BEC and BSC curves. This idea is verified by Fig. \ref{IaverursIe}, which illustrates the VND EXIT curves for length 3 repetition codes over the various channels. In the figure, $I_{VND,E}$ and $I_{VND,A}$ are the extrinsic MI and the average {\it a priori} MI of the VND, respectively. The BEC and BSC EXIT curves are obtained from \cite[Fig. 8]{land2005bounds}, while the AWGN channel curve is based on the channel noise variance \cite[Example 3]{ashikhmin2004extrinsic}. For the AWS$\alpha$SN channel, we adopt an EXIT curve generation method proposed for the non-Gaussian channel \cite{chen2013exit}. We remark that the method proposed by \cite{chen2013exit} generates the EXIT curves by using tedious and time-consuming simulations, and hence is not suitable to be used for the P-EXIT analysis for designing protograph codes. Observe from Fig. \ref{IaverursIe} that the EXIT curves for the AWS$\alpha$SN channel are very close to that of the AWGN channel, for both $\alpha=1.8$ (slightly impulsive noise) and $\alpha=1$ (strong impulsive noise).

By adjusting the noise variance of the AWGN channel, the mean squared error between the VND EXIT curves for the AWS$\alpha$SN channel and that for the AWGN channel can be as low as possible. Therefore, it is reasonable to use $L_{in}$ of the AWGN channel to approximate that of the AWS$\alpha$SN channel. This enables the derivation of the extrinsic MI of the VND for the AWS$\alpha$SN channel, given by
\begin{align} \nonumber \label{IEV}
I_{VND,E} & = I(X;L) = 1-E\left\lbrace \log_{2}\left( 1+e^{-L} \right)\right\rbrace \\ \nonumber
&\approx 1-\frac{1}{M}\sum^{M}_{n=1}\log_{2}\left( 1+e^{-L_{n,out,i}}  \right), \\
\end{align}
where $M$ denotes the number of $L_{out,i}$ samples. Here we consider that the all-zero codeword is transmitted. Correspondingly, the LLR output from the CND can be calculated using a "box-plus" operation and hence the extrinsic MI of the CND can be calculated similarly as \eqref{IEV}.

To further verify the effectiveness of the above proposed EXIT function, we compare the correspondingly obtained decoding thresholds with those computed by using the DE method \cite{mei2016finite}. Here, we consider the regular (3,6) and (4,8) LDPC codes and set the number of samples to be $M=30000$. The difference of the decoding thresholds obtained by using the two methods, denoted by $|\epsilon|$, is illustrated in Table \ref{Table 1} for different values of $\alpha$. Observe that the difference $|\epsilon|$ of the decoding thresholds obtained by using the two methods is at most 0.06 dB, over a wide range of $\alpha$, thus demonstrating the effectiveness of the proposed EXIT function for the AWS$\alpha$SN channel.

\begin{table}[t]\scriptsize
\centering
\caption{Difference of the decoding thresholds computed by using the two methods.}
\label{Table 1}
\begin{tabular}{|r|r|r|r|r|r|r|r|}
\hline
\multicolumn{4}{|r|}{Threshold of regular (3,6) LDPC code} & \multicolumn{4}{r|}{Threshold of regular (4,8) LDPC code} \\ \hline
$\alpha$        & EXIT(dB)        & DE(dB)       & $|\epsilon|$& $\alpha$        & EXIT(dB)       & DE(dB)       & $|\epsilon|$\\ \hline
1.9          & 1.34            & 1.33         & 0.01        & 1.9          & 1.77           & 1.75         & 0.02        \\ \hline
1.8          & 1.55            & 1.52         & 0.03        & 1.8          & 1.99           & 1.96         & 0.03        \\ \hline
1.7          & 1.73            & 1.69         & 0.04        & 1.7          & 2.19           & 2.15         & 0.04        \\ \hline
1.6          & 1.90            & 1.87         & 0.03        & 1.6          & 2.38           & 2.35         & 0.03        \\ \hline
1.5          & 2.09            & 2.05         & 0.04        & 1.5          & 2.64           & 2.59         & 0.05        \\ \hline
1.4          & 2.29            & 2.25         & 0.04        & 1.4          & 2.88           & 2.83         & 0.05        \\ \hline
1.3          & 2.51            & 2.47         & 0.04        & 1.3
& 3.10           & 3.06         & 0.04        \\ \hline
1.2          & 2.76            & 2.72         & 0.04        & 1.2
& 3.42           & 3.38         & 0.04        \\ \hline
1.1          & 3.03            & 2.98         & 0.05        & 1.1          & 3.76           & 3.71         & 0.05        \\ \hline
1.0          & 3.33             & 3.27        & 0.06        & 1.0            & 4.16           & 4.10         & 0.06        \\ \hline
\end{tabular}
\end{table}

\subsection{Novel Simulation-based P-EXIT Analysis for the AWS$\alpha$SN Channel}

Unlike the closed-form P-EXIT analysis based on the $J$ and $J^{-1}$ functions \cite{liva2007protograph} for the AWGN channels described in Section \textit{3.1.1}, the proposed novel simulation-based P-EXIT analysis is summarized as follows: \\
1) {\bf Initialization}. Given a $\frac{E_{b}}{N_{0}}$. According to \eqref{ebn0}, initialize a vector $\pmb{\gamma}=(\gamma_{0},\ldots,\gamma_{T-1})$ such that:
\begin{equation} \label{gamma}
\gamma_{t}=\sqrt{\frac{1}{4R_{c}C_{g}^{(\frac{2}{\alpha}-1)}(\frac{E_{b}}{N_{0}})_{t}}}.
\end{equation}
If $t$ is the puncture node, set $\gamma_{t}=0$. \\
2) {\bf VN to CN Renewal}. For $t\in \lbrace 0, \ldots ,T-1 \rbrace$ and $s\in \lbrace 0, \ldots ,S-1 \rbrace$, compute
\begin{equation} \label{LEij}
L^{(s,t)}_{out,s}=L^{t}_{ch}+\sum_{c=1}^{S}(b_{ct}-\delta_{cs})(L^{(c,t)}_{in,t}).
\end{equation}
If $c=s$, set $\delta_{cs}=1$. Otherwise, set $\delta_{cs}=0$. The MI of $L^{(s,t)}_{out,s}$, denoted by $I^{(s,t)}_{VND,E}$, can then be obtained by using \eqref{IEV}. \\
3) {\bf CN to VN Renewal}. For $t\in \lbrace 0, \ldots ,T-1 \rbrace$ and $s\in \lbrace 0, \ldots ,S-1 \rbrace$, compute

\begin{equation} \label{LCij}
L^{(s,t)}_{out,t}=\sum_{x=1}^{T} \boxplus (b_{sx}-\delta_{xt})(L^{(s,x)}_{in,t}).
\end{equation}
If $x=t$, set $\delta_{xt}=1$. Otherwise, set $\delta_{xt}=0$. The MI of $L^{(s,t)}_{out,t}$, denoted by $I^{(s,t)}_{CND,E}$, can then be obtained by using \eqref{IEV}. \\
4) {\bf A Posteriori MI Accumulation}. For $t\in \lbrace 0, \ldots ,T-1 \rbrace$, compute
\begin{equation} \label{Lmie}
L^{t}_{MIE}=L^{t}_{ch}+\sum_{c=1}^{S}(b_{ct})(L^{(c,t)}_{in,t}).
\end{equation}
The corresponding MI $I^{t}_{MIE}$ can be computed by measuring the LLRs $L^{t}_{MIE}$ as in \eqref{IEV}.
\\
5) {\bf Stopping criterion}. Return to Step 2 and carry on the iteration until $I^{t}_{MIE}=1$ for all $t$. The decoding threshold is the lowest value $\frac{E_{b}}{N_{0}}$ that meets the stopping criterion.

\subsection{Asymptotic Weight Distribution Analysis}

According to \cite{divsalar2009capacity}, the normalized logarithmic asymptotic weight distribution (AWE) $r(\delta)$ can be expressed as 
\begin{equation} \label{detection threshold}
S_{t}r(\delta)= \max_{\bm{\delta}_{t} : |\bm{\delta}_{t}| = S_{t}\delta}\left\lbrace  \sum_{i=1}^{T}a^{c}(\bm{\delta}_{i}) - \sum_{j=1}^{S}(d_{vj}-1)H(\delta_{j}) \right\rbrace,
\end{equation}
where $S_{t}$ is the number of transmitted VNs, $a^{c}(\bm{\delta}_{i})$ denotes the normalized logarithmic AWE for CN $c_{i}$ with normalized partial weight vector $\bm{\delta}_{t}$, $d_{vj}$ is the degree of the VN $vj$, and $H()$ is the binary entropy function \cite{divsalar2009capacity}. 

By exploting the AWD analysis, we can check whether the minimum distance of the code ensemble increases linearly with the codeword length or not.The second-zero crossing point of $r(\delta_{2c})$ is defined as the typical minimum distance ratio. If $\delta_{2c}>0$, then the minimum distance increases with the codeword length with a high probability, thus leading to a low error floor during decoding,

\section{Protograph LDPC Code Design}
The goal of designing a protograph LDPC code is to achieve a lower decoding threshold and at the same time the code ensemble is expected to have the linear minimum distance growth \cite{divsalar2009capacity}. Although AR4JA protograph codes \cite{divsalar2009capacity} optimized for the AWGN channel can preserve the linear minimum distance growth property, its decoding thresholds are 0.98 dB for $\alpha=1.8$ and 2.27 dB for $\alpha=1$, while those of the optimized code ensembles for the AWS$\alpha$SN channel \cite{dai2017exit} are 0.86 dB and 2.02 dB, respectively. Thus, the decoding thresholds achieved by the AR4JA code are higher than those achieved by the optimized irregular LDPC codes. Therefore, in the following, we propose the design of novel protograph LDPC codes for the AWS$\alpha$SN channel which achieve lower decoding thresholds than the AR4JA code.

Through testing over the AWS$\alpha$SN channel, it has been found that the VNs with degree-2 in the protograph lead to superior error rate performance when the G-SNR is low. However, as the number of the VNs with degree-2 increases, error floor may occur at the high G-SNR region  \cite{divsalar2009capacity}. Furthermore, we found that applying precoding to the protograph can decrease the decoding threshold \cite{divsalar2009capacity}. Based on the above observations, our proposed criteria for designing the protograph over the AWS$\alpha$SN channel are summarized as follows.
\par 1) Adjust the ratio of the VNs with degree-2 to obtain a protograph with the lowest decoding thresholds, and also guarantees the linear minimum distance growth property by using the AWD analysis. That is, if a normalized logarithmic AWD function begins at zero and returns back to zero after travelling to negtive values, the second-zero crossing is known as the typical minimum distance ratio. A positive value of the typical minimum distance ratio implies that there is a large chance that the code ensemble's minimum distance will increase linearly with the length of the code block, which leads to a low error floor for decoding.
\par 2) Apply precoding to the protograph obtained from Step 1 to further improve the decoding threshold.

Based on the above code design criteria, we start by constructing the rate$=$1/2 protograph LDPC codes, with $S=4$ and $T=7$. We then apply precoding and puncture the highest degree VNs. To reduce the search space, we search for the protograph subjected to the constraints: 
\begin{equation} 
   \begin{cases}
   2\leq\sum_{i=1}^{4}(b_{ij})\leq7, j\neq1,\\
   b_{i2}+b_{i6}\geq1, i=2,3,\\
   b_{ij}\in\lbrace0,1,2\rbrace.  
   \end{cases}
\end{equation}

\begin{figure}[t]
\centering
\includegraphics[height=0.2\columnwidth, width=0.51\columnwidth]{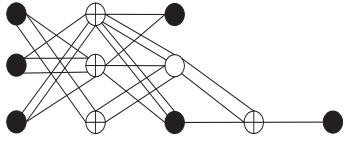}
\caption{Protograph of the designed rate-1/2 LDPC code.}
\label{proposed protograph code}
\end{figure}

In this way, we obtain the base matrix $\pmb{B}$ of the designed code, given by
\begin{equation}  \label{basematrixB}
 \pmb{B}=\left(
  \begin{array}{ccccccc}
    1 & 0 & 2 & 1 & 0 & 0 & 0\\
    0 & 1 & 2 & 2 & 0 & 1 & 2\\
    0 & 1 & 1 & 1 & 1 & 2 & 1\\
    0 & 0 & 2 & 0 & 2 & 0 & 0\\
  \end{array}
\right).
\end{equation}
Fig. \ref{proposed protograph code} illustrates the protograph structure of the designed code, where the white circle denotes the punctured node, and the black dots denote the normal nodes. The simulation-based P-EXIT analysis indicates that the correspondingly achieved decoding thresholds are 0.81 dB for $\alpha=1.8$, and 1.95 dB for $\alpha=1$. Furthermore, the second-zero crossing point of the design code is 0.006 according to the AWD analysis. This indicates that the constructed code ensembles for sure to have the linear minimum distance growth property\cite{divsalar2009capacity}.

\begin{figure}[h]
\centering
\includegraphics[height=0.8\columnwidth]{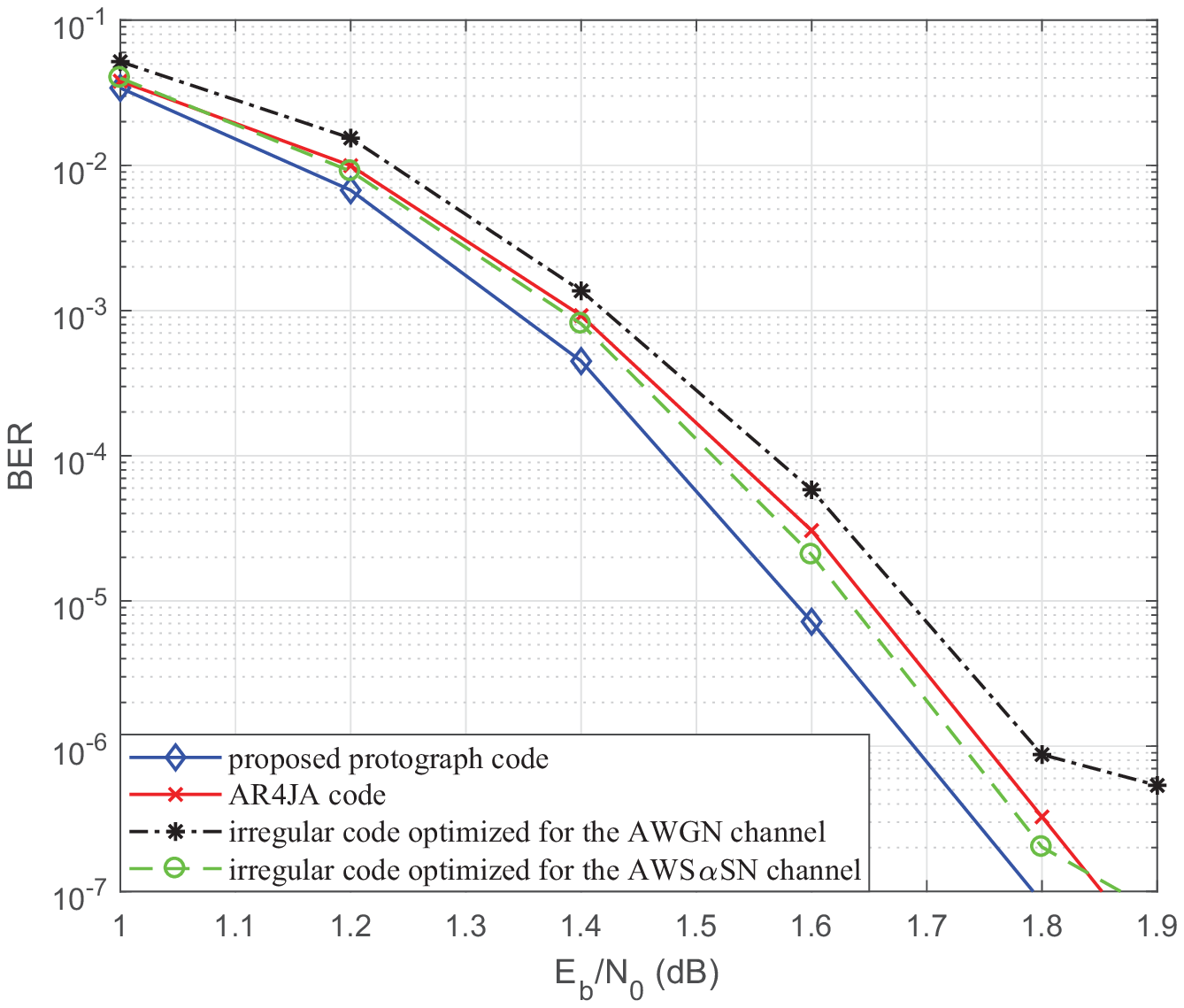}
\caption{BER comparison between the proposed protograph LDPC code and the prior art codes, for $N=8000$ and $\alpha=1.8$.}
\label{alpha=1.8}
\end{figure}

\begin{figure}[h]
\centering
\includegraphics[height=0.8\columnwidth]{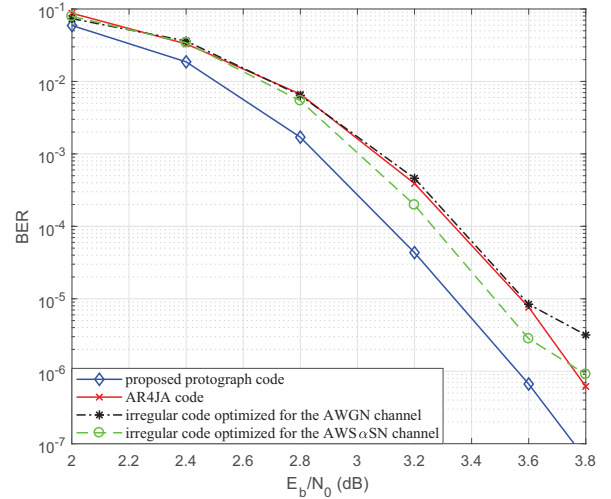}
\caption{BER comparison between the proposed protograph LDPC code and the prior art codes, for $N=8000$ and $\alpha=1$.}
\label{alpha=1}
\end{figure}

\section{Simulation Results}

In this section, we evaluate and compare the error rate performance of our designed protograph codes with various prior art LDPC codes proposed in the literature over the AWS$\alpha$SN channel, including the AR4JR code which is the protograph code optimized for the AWGN channel, the irregular LDPC codes optimized for the AWGN channel, as well as the irregular LDPC codes optimized for the AWS$\alpha$SN channel \cite{dai2017exit}. To meet up with the requirements of different applications, we construct protograph codes with different codeword lengths of N=8000, N=4000, and N=1000, respectively. For all the codes, the code rate is 1/2. With the help of the progressive-edge-growth (PEG) algorithm and circulant permutation \cite{hu2005regular}, the proposed protograph codes can be constructed easily.
In our simulations, we also take different levels of the noise impulsiveness of $\alpha = 1$ and $\alpha = 1.8$, respectively. The simulations have a maximum of 100 iterations per code block and are terminated after 100 block errors for each $E_{b}/N_{0}$.

Fig. \ref{alpha=1.8} and Fig. \ref{alpha=1} illustrate the bit-error rate (BER) comparison between the proposed code and the prior art codes when the codeword length is 8000 bits, for $\alpha=1.8$ and $\alpha=1$, respectively. From the figures, we observe that the proposed protograph code achieves the best performance over all the other codes, for both $\alpha=1.8$ and $\alpha=1$. In particular, both the irregular code optimized for the AWGN channel and the irregular code optimized for the AWS$\alpha$SN channel show an error floor at BER of around $10^{-6}$, while our proposed protograph codes and the AR4JA codes do not have an error floor. Furthermore, at BER = $10^{-5}$, our proposed codes outperform the AR4JA codes by nearly 0.1 dB with $\alpha=1.8$, and by 0.2 dB with $\alpha=1$. The above BER simulation results coincide with our simulation-based P-EXIT analysis and the AWD analysis.

\begin{figure}[t]
\centering
\includegraphics[height=0.8\columnwidth]{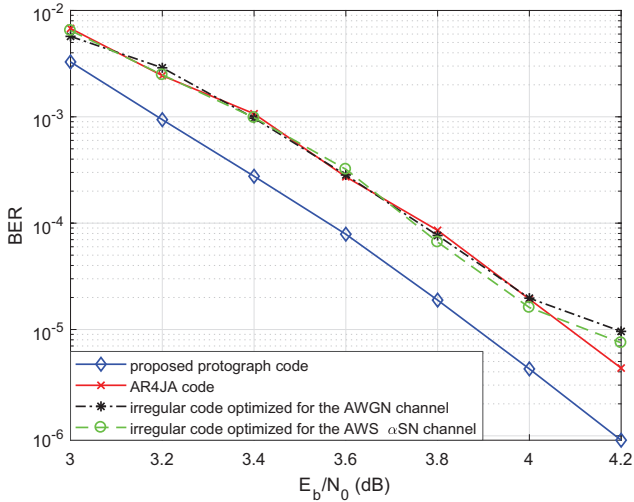}
\caption{BER comparison between the proposed protograph LDPC code and the prior art codes, for $N=4000$ and $\alpha=1$.}
\label{1Kalpha=1}
\end{figure}

\begin{figure}[h]
\centering
\includegraphics[height=0.8\columnwidth]{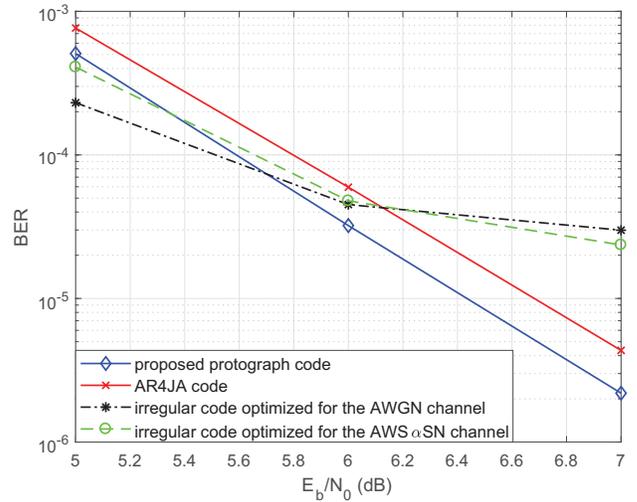}
\caption{BER comparison between the proposed protograph LDPC code and the prior art codes, for $N=1000$ and $\alpha=1$.}
\label{4Kalpha=1}
\end{figure}

In Fig. \ref{1Kalpha=1} and Fig. \ref{4Kalpha=1}, we further extend the BER comparison between the proposed codes and the prior art codes to two different codeword lengths of $N=4000$ and $N=1000$, for $\alpha=1$ only for the sake of simplicity. It is observed that the error floor of the the two types of irregular codes becomes higher with the decrease of the codeword length. Our proposed protograph codes achieve around 0.2 dB performance gains over the AR4JA codes for both $N=4000$ and $N=1000$. The above BER simulation results coincide with our simulation-based P-EXIT analysis and the AWD analysis, and demonstrate the potential of our proposed protograph LDPC codes for the AWS$\alpha$SN channel.

\section{Conclusion}

We have proposed a novel design of protograph codes for the AWS$\alpha$SN channel. Since the distribution of S$\alpha$S noise is non-Gaussian, the original P-EXIT analysis proposed for the AWGN channel cannot be applied to the AWS$\alpha$SN channel. Therefore, we have first derived an EXIT function for the AWS$\alpha$SN channel, based on which we proposed a novel simulation-based P-EXIT analysis to measure the decoding thresholds of the protograph codes designed for the AWS$\alpha$SN channel. We have further applied the AWD analysis to ensure that the code ensemble constructed to possess the linear minimum distance growth property. Both theoretical analysis and simulation results demonstrate that the proposed protograph code outperforms the prior art AR4JA code, the irregular codes optimized for the AWGN channel, as well as the irregular codes optimized for the AWS$\alpha$SN channel, for different codeword lengths and different levels of impulsiveness.

\section{Acknowledgment}
This work is supported by Singapore Agency of Science and Technology (A*Star) PSF research grant, Singapore Ministry of Education Academic Research Fund Tier 2 MOE2016-T2-2-054, and SUTD SRG grant SRLS15095.

\small
\bibliographystyle{IEEEtran}
\bibliography{refs}

\end{document}